\def\year{2022}\relax
\documentclass[letterpaper]{article} 
\usepackage{aaai22}  
\usepackage{times}  
\usepackage{helvet}  
\usepackage{courier}  
\usepackage[hyphens]{url}  
\usepackage{graphicx} 
\usepackage{natbib}  
\usepackage{caption} 
\DeclareCaptionStyle{ruled}{labelfont=normalfont,labelsep=colon,strut=off} 
\usepackage{algorithm}
\usepackage{algorithmic}
\usepackage[english]{babel}

\usepackage{xcolor}

\usepackage{newfloat}
\usepackage{listings}
\floatstyle{ruled}
\newfloat{listing}{tb}{lst}{}
\floatname{listing}{Listing}
\title{The Koo Dataset: An Indian Microblogging Platform With Global Ambitions}
\author{
    Amin Mekacher\textsuperscript{\rm 1}, Max Falkenberg\textsuperscript{\rm 1}, Andrea Baronchelli\textsuperscript{\rm 1, 2}\\
}
\affiliations{
    \textsuperscript{\rm 1}City University of London, Department of Mathematics, London EC1V 0HB, (UK)\\
    \textsuperscript{\rm 2}The Alan Turing Institute, British Library, London NW1 2DB, (UK)\\
    amin.mekacher@city.ac.uk, max.falkenberg@city.ac.uk, abaronchelli@turing.ac.uk
}

\usepackage{bibentry}

\begin{document}

\maketitle

\begin{abstract}
Increasingly, alternative platforms are playing a key role in the social media ecosystem. Koo, a microblogging platform based in India, has emerged as a major new social network hosting high profile politicians from several countries (India, Brazil, Nigeria) and many internationally renowned celebrities. This paper presents the largest publicly available Koo dataset, spanning from the platform's founding in early 2020 to September 2023, providing detailed metadata for 72M posts, 75M comments, 40M shares, 284M likes and 1.4M user profiles. Along with the release of the dataset, we provide an overview of the platform including a discussion of the news ecosystem on the platform, hashtag usage, and user engagement. Our results highlight the pivotal role that new platforms play in shaping online communities in emerging economies and the Global South, connecting local politicians and public figures with their followers. With Koo's ambition to become the town hall for diverse non-English speaking communities, our dataset offers new opportunities for studying social media beyond a Western context.
\end{abstract}

\section*{Introduction}

For many years, online social interactions have taken place primarily on well-established, US-based platforms such as Facebook, Twitter or Instagram. Consequently, significant research efforts have gone towards studying these platforms, for example considering their role in political campaigns \cite{punia2016_campaigns, McGregor2020_campaigns}, and the digital news ecosystem \cite{Buhl2016_news, Picard2014_news}. However, in recent years the social media landscape has evolved with the advent of the alt-tech ecosystem. These emerging platforms are becoming increasingly influential venues in the political communication ecosystem, attracting key right-wing politicians such as former US president Donald Trump and former Brazilian president Jair Bolsonaro \cite{allyn2022_truth, ribeiro2022_bolsonaro}, hosting major debates \cite{gilbert2023_rumble} and functioning as a political campaign ground prior to presidential elections \cite{leingang2023_truth}. 

Despite the growing importance of alt-tech platforms, research remains in its infancy in part due to a lack of publicly available data. This has motivated the release of datasets for many platforms including Gab \cite{Fair2019_gab} , BitChute \cite{Trujillo2022_bitchute}, Truth Social \cite{Gerard2023_truth} and Voat \cite{mekacher_2022voat}. However, these datasets overwhelmingly focus on platforms based in the West, catering largely to fringe, right-wing audiences. 
Launched in India in 2020, Koo has grown to become the second largest microblogging platform worldwide \cite{singh2022_platform}, and has attracted a number of political communities who have been critical of Twitter for censoring their discourse \cite{Bhat2021_alt}. Most notably, Koo has become a leading social platform used by the Bharatiya Janata Party (BJP), the governing political party in India. 

Given the dominance of the BJP on Koo, the platform appears to be similar to many other alt-tech platforms with a focus on right-wing politics. However, beyond India, Koo has successfully attracted a politically diverse set of users, including both supporters and opponents of former Nigerian president Buhari, and of Brazilian president Lula. This success highlights how Koo has the potential to move beyond the politically homogeneous model of most alt-tech platforms to become a major politically heterogeneous venue for both political and apolitical discussions across multiple countries, challenging the US social media hegemony. 

\textbf{Data Release.} In this work, we present the most extensive Koo dataset currently available, including 71.7M posts, 74.6M comments, 283.5M likes, 40.0M shares, and 1.4M user profiles. Our dataset spans from Koo's launch in early 2020 up until September 2023. Due to the ability to paginate through a user's complete timeline to access historical activity via the API, we are confident that our dataset provides a near-complete overview of the interactions carried out by users who are actively using the platform. 

\textbf{Relevance.} Our dataset provides researchers with the opportunity to study an alt-tech platform based in India which has attracted an international community of high-profile politicians and celebrities, primarily from India, Brazil and Nigeria. With BJP politicians heavily endorsing Koo, our dataset enables the study of political rhetoric through the analysis of the political content shared online. Previous research has shown that the BJP promote islamophobic \cite{Werleman2021_bjp} and populist \cite{Ammassari2022_bjp} rhetoric. It is therefore of interest to consider whether their claims on Koo aim to polarize opinions and stoke hate in their audience. Moreover, with many news outlets shared on Koo, the dataset will enable the study of political discourses through an analysis of the alignment between news media outlets' editorial line and the national political parties' ideologies \cite{Marques2019_newspaper, Barclay2014_newspaper}. Finally, the dataset can contribute to cross-platform comparisons of political rhetoric.

\textbf{Paper organization.} In this paper, we first explain Koo's structure before detailing its impact on the social media ecosystem. Second, we describe the method used to query posts, interactions and user profiles from Koo's public API before describing the structure of our dataset. Finally, we present results which provide an overview of the Koo platform, before discussing our results and the importance of our dataset in the conclusion.

\section*{What is Koo?}

Koo is a multi-lingual microblogging platform launched in early 2020 by Bombinate Technologies, a company based in Bangalore, India. Social interactions on Koo work in a very similar way as interactions on other microblogging platforms (e.g., Twitter): Users can create an account and then follow, or be followed by, other users. User profiles can be personalized with a profile-picture and additional information including a user description, personal title, and links to other social platforms. User profiles also display the account username and creation date.

Once logged in, users can submit public posts, called \textit{koos}. A user can like a post, comment on it, or share it (a \textit{rekoo}). Koo also provides the ability to translate a koo to many languages. Since March 2023, the platform has integrated ChatGPT in a bid to increase the number of users actively creating content for the platform \cite{dang2023_gpt}.

\begin{figure}[h!]
    \centering
    \includegraphics[scale=0.55]{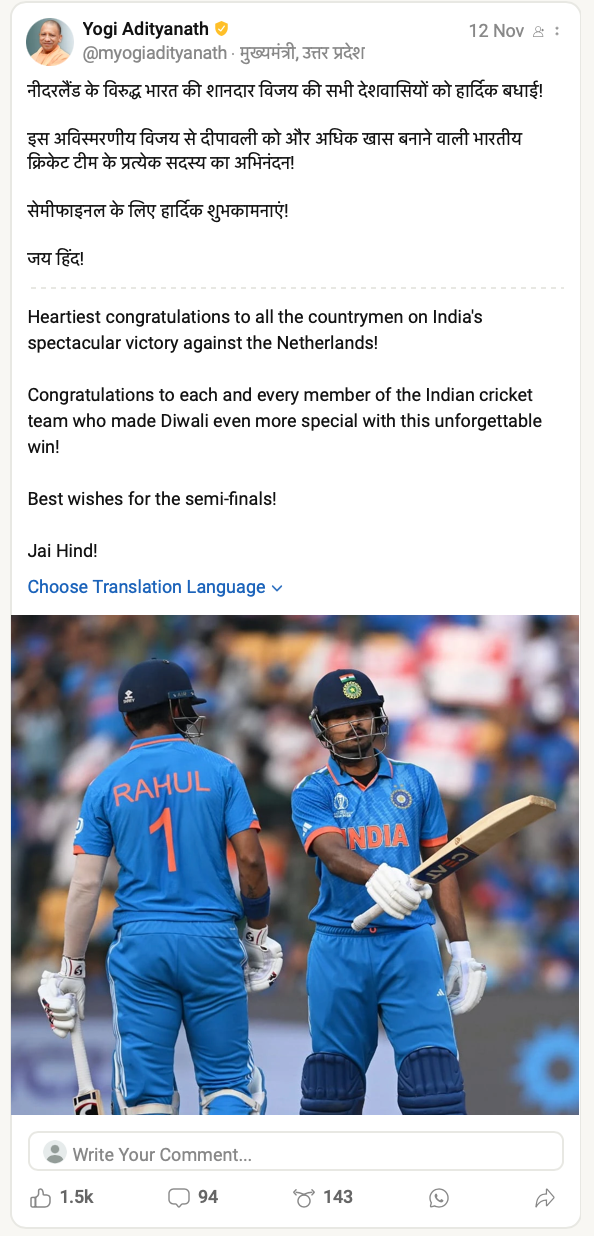}
    \caption{\textbf{An example of a koo.} The main panel includes the original post in Hindi, its translation in English and an image. The top panel provides information about the poster, including their user handle, profile picture, their self-declared title, the yellow tick of eminence (if applicable), and the post creation date. The bottom panel allows logged-in users to comment, share or like the post. Koo provides additional icons to share posts on other platforms.} 
    \label{fig:kohli_screenshot}
\end{figure}

Figure \ref{fig:kohli_screenshot} provides an example of a post on Koo. Posts have a 500 character limit and may include images and video. The panel under each post indicates how many times the koo has been commented, liked and shared by other users, and allows the viewer to share the koo on other platforms. The profile panel, located above the koo, displays the user's username and their self-indicated title. The yellow tick at the right of the username, known as the yellow tick of eminence, is granted by Koo administrators to accounts considered ``significant representative[s] of the Voices of the World'' \cite{httech2022_verif}. Other users can earn a green badge by self-verifying their account. 

\section*{The demographics of Koo}

Koo aims at becoming a central outlet for non-English speaking communities online by targeting countries where the uptake of traditional social media has been low \cite{phartiyal2021_india}. The platform has experienced three key surges in user registrations. First, in February 2021 Indian users joined Koo following an open conflict between Narendra Modi's cabinet and Twitter, after the government ordered Twitter to remove more than 1,000 accounts they alleged were responsible for spreading misinformation around the Indian farmers' protests \cite{phartiyal2021_india}. Many government officials and celebrities endorsed the platform in the wake of the conflict, including the chief minister of Uttar Pradesh Yogi Adityanath and cricket player Virat Kohli. Indian Prime Minister Narendra Modi has expressed his support for Koo but does not currently use the platform \cite{majumdar2021_koo}. Previous research has highlighted that about 80\% of the politicians who signed up on Koo are members of the BJP \cite{precog2022_bjp}. Consequently, Koo has been accused of supporting nationalist politics in India, despite its attempts to attract other political parties \cite{cornish2022_koo}.

Second, in June 2021, Twitter temporarily suspended then Nigerian president Muhammadu Buhari from the platform after he threatened his political opponents with violence. This led Buhari's government to instate a nationwide Twitter ban on June 5, 2021 \cite{paquette2021_nigeria}. Afterwards, Koo experienced a significant uptake among Nigerian government officials and civilians, both supporters and opponents of the Buhari regime, leading the platform to roll out accessibility for the main vernacular languages spoken in Nigeria \cite{abubakar2021_nigeria}. Koo quickly gained legitimacy within the country, becoming an official channel of communication for the government alongside Facebook and Instagram. Twitter was reinstated in Nigeria in January 2022, after they agreed to conditions set by the Nigerian government \cite{akinwotu2022_nigeria}, leading to a reduction in Koo's usage from 289,000 monthly active Nigerian users to 40,000 by September 2022 \cite{dosunmu2023_nigeria}.

Finally, in November 2022, shortly after Twitter was purchased by Elon Musk, Koo attracted a large Brazilian community following a Koo-related linguistic pun posted by Felipe Neto, a Brazilian YouTuber and online influencer \cite{gonzalez2022_brazil}. Consequently, Brazilian president Luiz Inácio Lula da Silva and many of his supporters joined Koo; Lula gained over 50,000 followers in less than 4 hours \cite{regina2022_lula} with Koo becoming the most downloaded app on the Apple App Store and the Google Play Store for a few days \cite{dcruze2022_brazil}. Koo is still on the lookout for new markets to expand into. The Philippines, Malaysia and Indonesia in particular are countries the platform is targeting \cite{httech2022_malaysia}.

Table \ref{tab:lang_prevalence} shows the total number of posts, comments, likes and shares made in the top 10 languages on the platform, as well as the percentage of posts made in each language. Hindi makes up about half of the total messages, highlighting the dominance of the Indian community on Koo. English, Portuguese and Nigerian English are the only non-Indian languages which have significant use on the platform. 

\begin{table*}[h!]
    \centering
    \begin{tabular}{cccccc}
        \textbf{Language} & \textbf{Posts} & \textbf{Ratio (\%)} & \textbf{Comments} & \textbf{Likes} & \textbf{Shares}\\
        \hline
         Hindi & 35,074,082 & 48.9 & 33,762,199 & 141,851,555 & 22,710,161 \\
         English & 21,341,502 & 29.8 & 15,364,430 & 59,405,972 & 10,745,495 \\
         Portuguese & 4,933,641 & 6.88 & 9,747,183 & 51,323,645 & 1,521,073  \\
         Telugu & 2,127,279 & 2.97 & 2,218,213 & 1,975,845 & 104,001 \\
         Kannada & 1,889,473 & 2.63 & 3,700,669 & 5,972,733 & 859,069  \\
         Marathi & 1,286,933 & 1.79 & 913,997 & 5,237,631 & 128,886  \\
         Tamil & 1,268,948 & 1.77 & 376,721 & 1,262,414 & 82,940  \\
         Bengali & 1,194,315 & 1.67 & 1,031,905 & 2,912,268 & 203,597  \\
         Gujarati & 1,014,200 & 1.41 & 504,975 & 2,991,499 & 103,317  \\
         Nigerian English & 610,568 & 0.85 & 1,076,604 & 1,893,175 & 571,813  \\
         \hline
    \end{tabular}
    \caption{The top 10 languages used on Koo. The ratio indicates the percentage of the total number of posts written in each language. Columns indicate, for each language, the number of comments, likes and shares associated with the language.}
    \label{tab:lang_prevalence}
\end{table*}

Focusing on high-profile Koo users, Table \ref{tab:badge_lang} shows the fraction of users with a green (verified) or yellow (account of eminence) badge for the 10 largest linguistic communities on Koo. Over a quarter of Portuguese-speaking users have received a green badge, the largest percentage for any linguistic community, indicating heavy use of Koo's self-verification feature. In contrast, only 0.33\% of Nigerian English accounts are self-verified. Hindi-speaking users also display a high level of adoption for the self-verification process. For each linguistic community, only a small fraction of accounts are awarded the yellow tick of eminence, with a slight bias favouring Indian communities. 

\begin{table*}[h!]
    \centering
    \begin{tabular}{cccc}
        \textbf{Language} & \textbf{Self-Verified (\%)} & \textbf{Accounts of Eminence (\%)} & \textbf{User Profiles}\\
        \hline
         Hindi & 10.8 & 0.79 & 552,941\\
         English & 7.33 & 1.37 & 212,244\\
         Portuguese & 27.96 & 0.5 & 178,471\\
         Telugu & 4.05 & 0.35 & 16,265 \\
         Kannada & 2.95 & 0.93 & 25,547\\
         Marathi & 5.81 & 0.63 & 11,226\\
         Tamil & 5.3 & 1.15 & 5,021\\
         Bengali & 5.66 & 0.34 & 14,619\\
         Gujarati & 5.51 & 0.58 & 9,931\\
         Nigerian English & 0.33 & 0.31 & 17,184\\
         \hline
    \end{tabular}
    \caption{Ratio of self-verified accounts and those with a yellow tick of eminence for the top 10 linguistic communities on Koo. The table also indicates the number of user profiles provided for each linguistic community.}
    \label{tab:badge_lang}
\end{table*}

\begin{figure}[h!]
    \centering
    \includegraphics[width=\columnwidth]{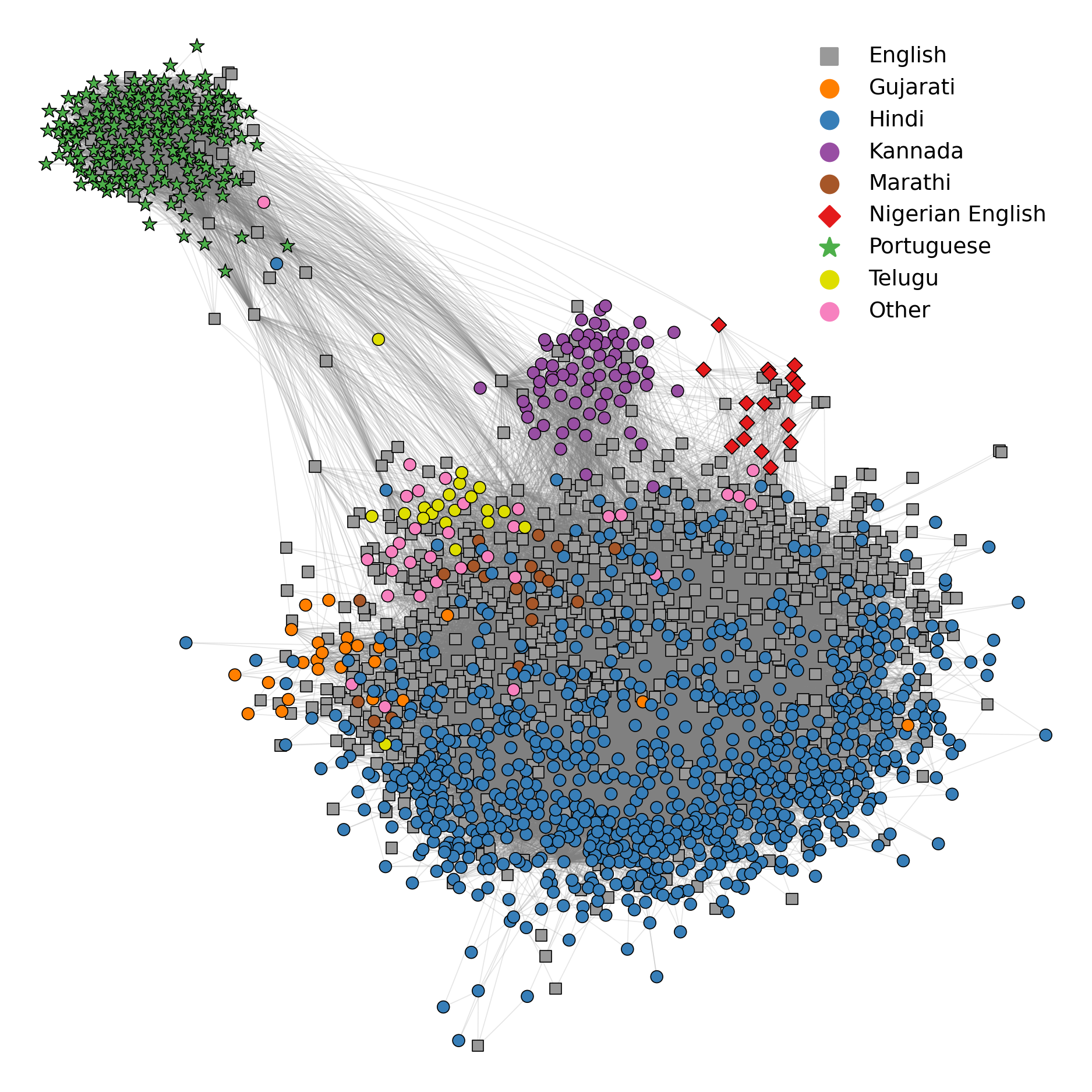}
    \caption{\textbf{Co-occurrence network of accounts of eminence.} Two eminent users are connected by an edge if at least 50 accounts on Koo interact with both of them. Nodes are coloured according to modal account language. Node shapes differentiate Indian and non-Indian languages.}
    \label{fig:netviz}
\end{figure}

Studies have shown that two accounts who share a common set of social media interactions typically share similar political or ideological views \cite{falkenberg2022growing, falkenberg2023_polarization}. To map these connections on Koo, we construct the co-occurrence network of Koo users with the account of eminence badge following the methodology outlined in \cite{falkenberg2023_polarization}, see Fig.~\ref{fig:netviz}. Each node in the co-occurence network corresponds to a single eminent user, coloured according to their modal post language. Two eminent users are connected by an edge if their posts are liked by at least 50 common users. For visual clarity, we only show the giant connected component, and eminent users with edges to at least two other users. 

The network shows that clusters are strongly influenced by linguistic factors: prominent users are more likely to be part of a similar social circle on Koo if they predominantly use the same language on the platform. Aside from the isolated Portuguese-speaking cluster and the Nigerian English speaking users, eminent users speaking Kannada, Telugu and Gujarati are also slightly disconnected from the English/Hindi dominant community. However, there is little evidence of structural polarization within linguistic communities on Koo, unlike on Twitter \cite{falkenberg2023_polarization}.

\section*{Data Collection}

\textbf{Koo's API.} Koo provides a public API, which can be accessed by sending a GET request to \textbf{https://www.kooapp.com/api/}. All data included in this dataset was retrieved by querying the user profile endpoint: knowing a user ID or account handle, the API provides specific endpoints to retrieve their posts, comments, shares and likes.

\textbf{User Profile Collection.} In order to collect a user's activity, we first need to retrieve their profile to get their user ID. We started with a short list of manually collected high-profile users, including prominent politicians, who are active on Koo. Through their interactions on Koo (see below) we identify any user they have interacted with. We query the user profile of these accounts, and repeat this process iteratively, collecting more users through a snowball sampling method until we retrieve no new user profiles. A limitation of this data acquisition strategy is that we do not capture lurkers on Koo (users who use Koo passively). However, we do capture all the interactions for every user who interacted with any account in the giant connected component on Koo. The user profiles provided by the API include information on a user's username, display name, profile description and manually entered title.

\textbf{User Activity.} Given a user ID, Koo's API provides specialized endpoints to query the user's posts, comments, shares and likes. The endpoints are respectively \textbf{profile/createdKus}, \textbf{profile/commentsKus}, \textbf{profile/rekooKus} and \textbf{profile/likedKus}, with the user ID provided as a query parameter. We paginate through the results to ensure that all of a user's posts and interactions are collected.

\textbf{Fair Principles.} The data released in this paper align with the FAIR guiding principles for scientific data:

\begin{itemize}
    \item \textit{Findable: } We assign a unique constant digital object identifier (DOI) to our dataset.
    \item \textit{Accessible: } Our dataset is openly accessible. 
    \item \textit{Interoperable: } Releasing the data with JSON files ensures that it can easily be read with various programming languages and operating systems. 
    \item 
    \textit{Reusable: } We provide the data schema in the \textit{Data Description} Section , to ensure that the purpose of each data field is clear. 
\end{itemize}

\textbf{Ethical Considerations.} The data released with this paper was collected using Koo's public API, in accordance with the platforms' Terms of Service. Koo does not offer the possibility for users to make their account private, therefore the data we are releasing contains publicly accessible information only. We do not release metadata related to geolocation. All data released are available publicly and results are aggregated across accounts. To ensure that our data collection pipeline and data sharing complies with the relevant regulations, we have completed a Data Protection Impact Assessment which has been approved by our institution. 

\section*{Data Description}

We now present the structure of the dataset, available on a public repository \cite{mekacher2024_dataset}. We release the metadata related to each type of user activity, as well as the user profiles, in separate JSON files. Table \ref{tab:metadata} indicates, for each user activity, the fields included in the JSON schema, including the data type of each field and a short description. For user convenience, we provide language-specific files for each interaction type. In total, we release 34 files for comments, shares and likes, and 43 files for the posts. This difference is due to small linguistic communities with fewer than 5 posts and no other interaction types.
\begin{table}[h!]
    \centering
    \scalebox{0.85}{
    \begin{tabular}{ccc} 
         \textbf{Key}&  \textbf{Type}& \textbf{Description} \\
         \hline
         \multicolumn{3}{c}{\textbf{language\_code\_posts.json} (43 files)} \\ 
         \hline
         id&  string&  ID of the post\\ 
         creatorId&  string&  ID of the user who created the post \\ 
         title&  string&  Content of the post \\ 
         createdAt&  int&  UNIX timestamp of the post\\ 
         handle&  string&  Handle of the user who created the post \\
 \hline
 \hline
 \multicolumn{3}{c}{\textbf{language\_code\_likes.json} (34 files)} \\
 \hline
 id& string& ID of the liked post\\
 creatorId& string& ID of the user who created the post\\
 createdAt& int & UNIX timestamp of the like\\
 handle& string& Handle of the user who posted\\
 liker\_id& string& ID of the user who liked\\
 \hline
 \hline
 \multicolumn{3}{c}{\textbf{language\_code\_comments.json} (34 files)} \\
 \hline
 id& string&ID of the commented post \\
creatorId& string&ID of the user who created the post \\
 title& string&Content of the comment \\
 createdAt& int&UNIX timestamp of the comment \\
 handle& string&Handle of the user who posted \\
 commenter\_id& string&ID of the user who commented \\
 \hline
 \hline
 \multicolumn{3}{c}{\textbf{language\_code\_shares.json} (34 files)} \\
 \hline
 id& string& ID of the shared post\\
 creatorId& string& ID of the user who created the post\\
 createdAt & int& UNIX timestamp of the share\\
 handle & string & Handle of the user who posted\\
 sharer\_id & string& ID of the user who shared\\
 \hline
 \hline
 \multicolumn{3}{c}{\textbf{koo\_users.json} (1 file)} \\
 \hline
 id& string& ID of the user\\
 handle & string & Handle of the user \\
 title & string & Self-given title \\
 description & string & Profile description \\
 createdAt & int & Timestamp of the account creation \\
 \hline
    \end{tabular}}
    \caption{Description of the metadata in each data file.}
    \label{tab:metadata}
\end{table}

\section*{Results}

In this section, we provide a quantitative overview of Koo including (1) an analysis of user engagement over time (2) an analysis of Koo's news media ecosystems, and (3) an analysis of user content. These results offer a more in-depth description of Koo's growth within the digital platform ecosystem and its potential to harbour a unique online community.

\begin{figure*}[h!]
    \centering
    \includegraphics[scale=0.8]{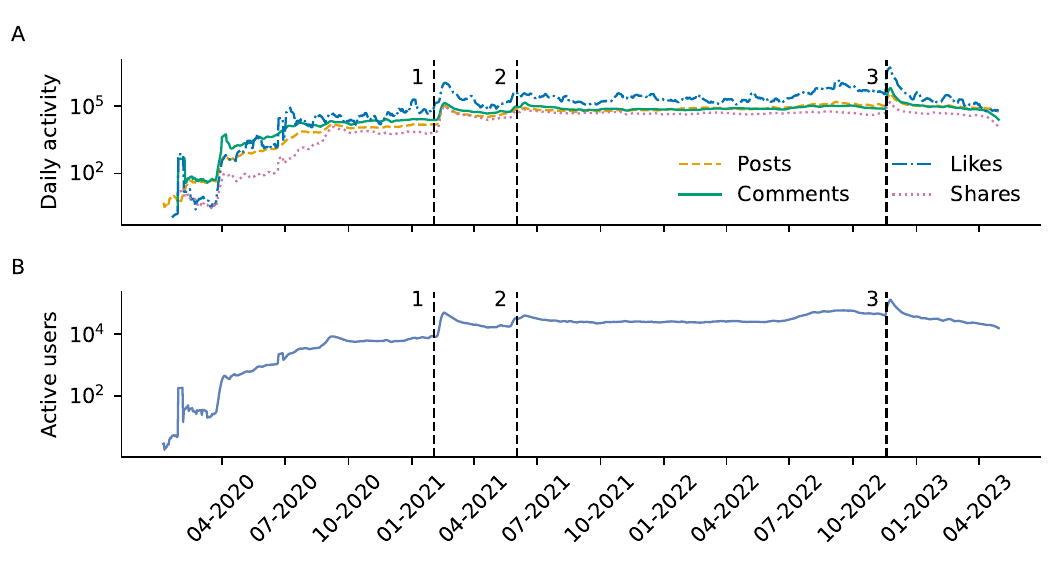}
    \caption{\textbf{Daily activity and number of active users.} A) 7-day moving window average of the amount of content (posts, comments, likes and shares) posted on Koo. B) 7-day moving average of the number of active users on a given day. A user is considered active if they created a new post or if they commented, shared or liked an existing post. The dashed lines indicate the events that led to the major collective migrations on Koo, namely 1) the Farmers' Protest in India; 2) Twitter getting banned in Nigeria and 3) Elon Musk's purchasing Twitter and the subsequent Brazilian migration.}
    \label{fig:activity}
\end{figure*}

\textbf{User retention.} Previous studies have shown that, although alternative platforms are successful at attracting a large number of user registrations, many newcomers do not remain active on the platform and become idle within days of registering \cite{thiel2021gettr,mekacher2023_gettr}. Figure \ref{fig:activity}A shows the number of posts, comments, shares and likes, recorded on a daily basis on Koo. Figure \ref{fig:activity}B shows the daily number of active users over time. 

After steady growth in 2020, the platform experienced a jump in activity in February 2021, attributed to BJP officials signing up on Koo with their followers. The daily activity count starts plateauing afterwards, but does not experience any substantial decrease. We see a second major increase in activity in November 2022 after Brazilian celebrities endorsed Koo, although activity levels quickly return to levels observed before the Brazilian migration. User engagement and activity does not fall substantially after any of the spikes in user registrations. This result may indicate low retention amongst users who joined Koo during the major collective migrations, a pattern that has been observed in other alt-tech outlets \cite{mekacher2023_gettr, mekacher_2022voat}. 

\textbf{News ecosystem.} Next, we look at news media URLs shared on Koo. Previous studies have highlighted the emergence of online ecosystems where users are only exposed to a limited selection of sources, often due to information-filtering mechanisms \cite{Sullivan2020_epistemic, Handfield2023_epistemic}. This phenomenon is commonly referred to as an \textit{epistemic bubble} and has been implicated in online radicalisation processes \cite{Nguyen2018_epistemic} and populist discourses \cite{anderson2021_epistemic}. The question of news diversity is of particular interest in the case of an Indian social platform that aims to host a wide range of cultural backgrounds. Previous studies have underlined the ethical issues that arise when religious events are covered in a controversial way by mainstream outlets that do not promote media secularism \cite{Ramesh2022_religion, thomas2021_religion}. 

\begin{figure}[h!]
    \centering
    \includegraphics[width = \linewidth]{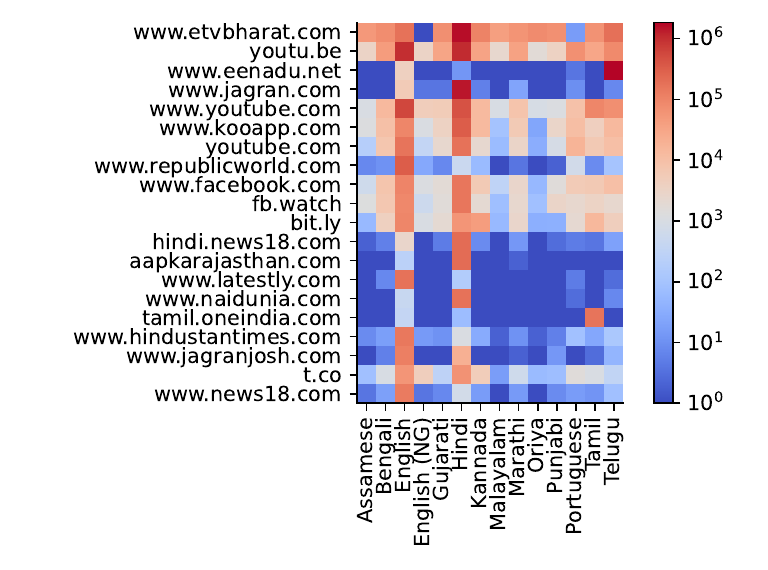}
    \caption{\textbf{Top-shared web domains and their prevalence in the dominant linguistic communities.} Number of links leading to a web domain shared by the top-10 linguistic communities on Koo. The top-20 shared domains are shown.}
    \label{fig:link_prevalence}
\end{figure}

To measure the news ecosystem for each linguistic community, we separate posts into languages for the news media analysis. This approach allows us to map the news ecosystem shared among users who predominantly use the same language on the platform. Comparing the results across languages demonstrates an outlet's ability to seep through language barriers. Figure \ref{fig:link_prevalence} shows the top 20 online domains shared on Koo and their respective prevalence within the major linguistic communities on the platform. We notice a strong tendency for Hindi news outlets to be part of the top domains on Koo, as Hindi speakers represent almost half of Koo's user base. However, some websites are also broadly shared within other linguistic communities. This is the case of \textit{ETV Bharat}, an Indian news channel that is available in 11 major Indian languages. Social media platforms, such as YouTube and Facebook, are also widely shared across linguistic communities.

Some news outlets are found to be very popular amongst smaller Indian linguistic communities. This is the case for \textit{Eenadu}, the most prominent daily newspaper in Telugu. Along with \textit{Sakshi}, it reaches more than 70\% of the Telugu-speaking audience \cite{mom2021_india}. For the Tamil-speaking community, the lesser-known news media \textit{OneIndia} is widely shared. \textit{OneIndia} has been very prolific on Koo and has garnered a large following by systematically sharing links to its articles, thereby outranking some of the better known Tamil newspapers such as \textit{Dinakaran}.

Interestingly, when looking at Nigerian English speakers, the most shared outlet is \textit{Peoples Gazette}, an online newspaper launched in 2020 that has led several investigations about cases of corruption in Muhammadu Buhari's government \cite{adenekan2022_gazette}. The second most shared news outlet among Nigerian English speakers (exc. YouTube and Telegram) is \textit{The Punch}, the most widely read newspaper in Nigeria, which is also highly critical of Buhari's politics \cite{times2018_punch}. These findings suggest that the Nigerian government was also followed by its dissenters when it migrated to Koo in June 2021 and has a less prominent influence on the relevant news ecosystem than their political opponents. 

Portuguese speakers widely share links to social media platforms, with Instagram, YouTube and Koo links being the most shared websites. These are followed by \textit{G1 Globo}, one of the most popular news outlets in Brazil. Similar to the other major newspapers in Brazil, \textit{Grupo Globo}, the conglomerate that owns \textit{G1 Globo}, was highly critical of Jair Bolsonaro's political stances \cite{newman2021_reuters}. However, the network has also been accused of delegitimizing Lula and other leaders of the Workers' Party, such as former president Dilma Rousseff \cite{vanDijk2017_globo}. In a similar fashion as for Nigeria, these results suggest that the Brazilian community on Koo share news media content which is primarily antagonistic to the current political regime. However, the Brazilian migration also included media actors that are supportive of Lula's regime. The second most shared news platfom within the Brazilian community is \textit{Diario do Centro do Mundo}, a left-leaning digital news outlet criticised as ``charlatans hired by the Workers' Party'' \cite{paiva2023_dcm}. 

\begin{figure}[h!]
    \centering
    \includegraphics[width=\linewidth]{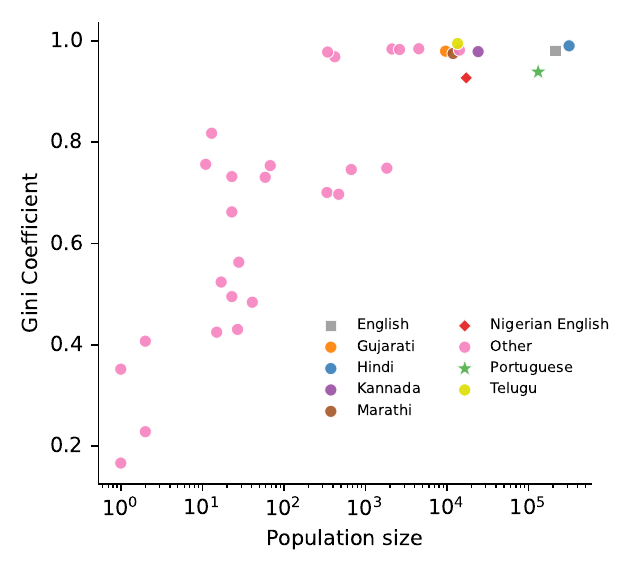}
    \caption{\textbf{Media plurality across linguistic communities on Koo.} The Gini coefficient for the news media web domains shared by each linguistic community, plotted against the population size of each linguistic community. A Gini coefficient close to 1 highlights a monopoly held by a single news source, whereas a Gini coefficient close to 0 indicates a more diverse link-sharing ecosystem.}
    \label{fig:gini_sources}
\end{figure}

The dominance of specific news outlets within linguistic communities, particularly in India, suggests that Koo may not harbour, or be conducive to, a social environment that cultivates media pluralism. Media pluralism has often been defined as a hallmark of a healthy democracy \cite{Keller2011_pluralism}, while also being an important tool for tackling misleading news \cite{Joris2020_news}. To measure news diversity, we compute the Gini coefficient for the set of web domains shared by each linguistic community on Koo. A Gini coefficient of 1 indicates that only one news source is being shared within the community, whereas a coefficient of 0 would represent perfect equality across all news sources. Figure \ref{fig:gini_sources} shows the Gini coefficient for each linguistic community, plotted against its population size. The figure shows that larger linguist communities on Koo tend to have a Gini coefficient close to 1. This further highlights the dominance of individual news media outlets in each of these communities. 

\textbf{Hashtags analysis.} Next, we look at the conversations being hosted on Koo. Previous studies focused on the US alt-tech ecosystem have shown that populist themes are widely discussed in culturally homogeneous communities \cite{allchorn2023_farright}. To identify the most salient topics users discuss, we perform an analysis of the hashtags shared by the major linguistic communities on Koo. Often used to bring visibility to a specific topic, hashtags have been shown to play an instrumental role on several social platforms, both in political campaigning and to rally citizens to activist movements. These social phenomena have been extensively studied in India \cite{losh2014_activism, panda2020_politics}, Nigeria \cite{opeibi2019_politics, akpojivi2019_activism} and Brazil \cite{almeida2020_activism, soares2021_politics}.

Table \ref{tab:hashtags_table} shows the 20 most shared hashtags for the Hindi, English, Nigerian English and Portuguese speaking Koo communities. For Hindi and English speakers, we note that the most popular hashtags gravitate around Indian matters (\#kooforindia, \#india), suggesting that the English-speaking community are largely based in India. Most Hindi and English hashtags relate to national politics and economic matters (\#narendramodi, \#HindiNews, \#Budget2022), which further highlights that Koo has succeeeded in hosting a sustainable conversation within India. However, we also find apolitical hashtags, particularly related to Cricket in India (\#Cricket, \#iplauction2022). Finally, this analysis highlights interest in Rampal Singh Jatain (\#Sat\_Bhakti\_Sandesh, \#SantRampalJiMaharaj), an Indian religious cult leader who has been previously arrested for murder \cite{chandra2018_rampal}. He signed up on Koo in August 2020 and garnered more than 330k followers.

The Nigerian English-speaking community prominently uses hashtags related to the Twitter ban and the migration to Koo in June 2021 (\#letskoo, \#bantwitter, \#nigeriatwitterban). Furthermore, many hashtags also mention locations in the country, such as cities (\#kano, \#kaduna, \#lagos) or regions (\#arewa). These results show that the Twitter ban remained a major topic of debate, and that Nigerian users attempted to rebuild their social network on Koo by connecting with peers located in the same region of the country. National politics are also discussed (\#Buhari, \#buharitrain, \#OsinbajoDay) and highlight the presence of both pro- and anti-Buhari cohorts on the platform. The \#OsinbajoDay hashtag refers to then Vice President Yemi Osinbajo, hailed by grassroots campaigners as the right candidate to take over ``the good works already started by Muhammadu Buhari'' \cite{oluwafemi2021_osinbajo}. In contrast, the \#buharitrain hashtag is used on social media to criticize president Buhari's railway project between the capital city of Abuja and its airport, with the investment considered a failure \cite{bd2023_train}. 

The hashtags used by the Brazilian community highlight a desire from users to connect with their peers on Koo (\#koobrasil and \#sdv, Portuguese slang for ``follow me back'' on social platforms). This is further suggested by the prominence of hashtags related to Elon Musk's purchase of Twitter (\#elonmusktwitter, \#layoffs, \#riptwitter). The initial migration indeed occupies a significant space in the Brazilian conversation, with other prominent hashtags related to the linguistic pun that triggered the migration to Koo (\#brasilnokoo, \#tomandonokoo) \cite{gonzalez2022_brazil}. Unlike the Indian and Nigerian communities, none of the most prevalent hashtags in the Brazilian community are related to contemporary political topics. Instead, mentions of the 2022 FIFA World Cup (\#copadomundo2022, \#fifaworldcup, \#vaibrasil) and general conversations (\#memes, \#humor, \#art) are more common. These results highlight that the Brazilian migration, unlike the Indian and the Nigerian ones, was not initially politically motivated. 

Overall, the hashtags indicate that the main communities on Koo are all involved in conversations relating to the events which triggered their communities' respective migrations to Koo. This highlights the impact that social media deplatforming can have on online discourses \cite{mekacher2023_gettr}. The stark difference in the prominence of political hashtags between the Brazilian community and the Nigerian / Indian communities underlines that the initial motivations leading to a platform migration can be both political and apolitical, and can  heavily impact the dominant discourses that subsequently emerge on alt-tech platforms.

\begin{table*}[h!]
\centering
\scalebox{0.92}{
\begin{tabular}{|ll|ll|ll|ll|}
\hline
\multicolumn{2}{|c|}{\textbf{Hindi}} & \multicolumn{2}{c|}{\textbf{English}} & \multicolumn{2}{c|}{\textbf{Nigerian English}} & \multicolumn{2}{c|}{\textbf{Portuguese}} \\ \hline
\textbf{Hashtag}    & \textbf{Ratio (\%)} & \textbf{Hashtag}     & \textbf{Ratio (\%)} & \textbf{Hashtag}      & \textbf{Ratio (\%)}    & \textbf{Hashtag}     & \textbf{Ratio (\%)}    \\ \hline
koooftheday         & 0.77           & koooftheday          & 0.62           & nigeria               & 5.41                   & koobrasil            & 4.46              \\
comedy              & 0.72           & india                & 0.37           & letskoo               & 3.38                   & brasilnokoo          & 3.78              \\
memes               & 0.71           & memes                & 0.32           & Nigeria               & 2.34                   & koo                  & 3.77              \\
memeoftheday        & 0.71           & India                & 0.31           & bantwitter            & 1.42                   & copadomundo2022      & 3.61              \\
dailymemes          & 0.42           & koo                  & 0.31           & kano                  & 1.32                   & brasil               & 2.94              \\
Budgetmemes         & 0.37           & comedy               & 0.30           & kanoconnect           & 1.14                   & tomandonokoo         & 2.56              \\
SaintRampalJi       & 0.31           & kooforindia          & 0.29           & koo                   & 1.04                   & meme                 & 1.57              \\
HindiNews           & 0.29           & memeoftheday         & 0.28           & kaduna                & 1.01                   & koonobrasil          & 1.42              \\
koo                 & 0.25           & narendramodi         & 0.27           & lagos                 & 0.95                   & riptwitter           & 1.30              \\
SantRampalJiMaharaj & 0.25           & motivation           & 0.25           & arewapeeps            & 0.87                   & memes                & 1.13              \\
Sat\_Bhakti\_Sandesh*   & 0.24           & dailymemes           & 0.25           & nigeriatwitterban     & 0.82                   & elonmusktwitter      & 0.79              \\
india               & 0.24           & trending             & 0.23           & kooyouropinion        & 0.81                   & humor                & 0.78              \\
MpNews              & 0.20           & Budget2022           & 0.20           & june12                & 0.80                   & layoffs              & 0.77              \\
entertainment       & 0.20           & Budgetmemes          & 0.19           & arewakooconnect       & 0.78                   & koobr                & 0.75              \\
Navabharat          & 0.19           & Cricket              & 0.18           & ifollowback           & 0.67                   & fifaworldcup         & 0.75              \\
Nan                 & 0.19           & news                 & 0.18           & OsinbajoDay           & 0.67                   & copadomundo          & 0.64              \\
kooforindia         & 0.18           & COVID19              & 0.18           & covid19               & 0.66                   & vaibrasil            & 0.56              \\
iplauction2022      & 0.18           & SaintRampalJi        & 0.17           & Buhari                & 0.64                   & Brasil               & 0.54              \\
narendramodi        & 0.18           & SantRampalJiMaharaj  & 0.17           & Arewa                 & 0.64                   & sdv                  & 0.54              \\
kookiyakya          & 0.18           & life                 & 0.15           & buharitrain           & 0.63                   & art                  & 0.53              \\ \hline
\end{tabular}}
\caption{\textbf{Top hashtags used on Koo.} Prevalence of the top 20 hashtags used by the major linguistic communities on Koo. The percentage indicates the percentage of usage of the respective hashtag within the considered linguistic community. The hashtag indicated with the asterisk (*) is translated from Hindi.}
\label{tab:hashtags_table}
\end{table*}

\section*{Conclusion}

In this work, we release and present a Koo dataset totalling over 72M posts, 399M user interactions, and 1.4M user profiles. The whole dataset was collected via Koo's undocumented public API. 

Our analysis highlights the growth of a multi-lingual, cross-country online ecosystem, with collective migrations to the platform often spearheaded by influential politicians. Many narratives on the platform relate to the events which triggered the initial migrations to Koo, resulting in prominent debates on the national political landscape of India and Nigeria, and to a lesser extent Brazil. For India, Koo content is heavily biased in support of the ruling BJP party. In contrast, both the Nigerian and Brazilian communities on Koo share pro- and anti-Government content, although the Brazilian Koo discussion is less political than the others.

These collective migrations highlight how the success of an alt-tech fringe platform can be intimately linked to the decisions made by mainstream platforms and to platform- or regulator-driven deplatforming events (e.g., because Twitter banned users, or because regulators banned Twitter) \cite{mekacher2023_gettr}. Catering to these communities, Koo has become a prominent venue for political and social debates for several communities. However, despite a lack of news media and content diversity, and unlike most alt-tech fringe platforms, Koo has had success in attracting a politically heterogeneous user-base beyond India in both Nigeria and Brazil. This sets Koo apart from other emerging platforms based outside the US - especially those in China and Russia where politicians have emphasised the need for ``Internet sovereignty'' \cite{broderick2023_russia} - in that their ambition is to expand beyond their home jurisdiction. Consequently, a platform like Koo has the potential to challenge the current dominance of US-based social media platforms, a change which may have important consequences for social media regulation.  

We anticipate that this dataset will support further research investigating digital ecosystems based outside the US, and how they impact political campaigning and news sharing. This is particularly important given upcoming elections in several major democracies, including India, in 2024 \cite{hsu2024_election}. Given the well-documented risks that social media can be used to promote political disinformation \cite{hsu2024_election}, and that Koo attracted some users who were deplatformed from its mainstream competitors, the platform may face challenges related to content moderation, balancing its free-speech agenda with its aim to prevent the spread and amplification of hate speech. 

Finally, given Koo's linguistic diversity, we hope that our dataset will motivate researchers to further develop computational tools for studying social media narratives and news media sharing in underrepresented languages. This work will be crucial to enable a robust analysis of the online news media ecosystem, misinformation, and its potential impact on political campaigns across diverse markets. 

\bibliography{aaai22}

\end{document}